\documentclass{article}
\usepackage{caption}
\usepackage{subcaption}
\usepackage{authblk}
\usepackage{amsmath}
\usepackage{graphicx}
\usepackage{amssymb}

\usepackage{amsmath}
\usepackage{bm}
\usepackage{siunitx} 

\begin{document}

\title{High-Fidelity Description of Platelet Deformation Using a Neural Operator}

\author[1]{Marco Laudato\footnote{Corresponding authors; laudato@kth.se, manzari@kth.se}}
\author[2]{Luca Manzari$^*$}
\author[3]{Khemraj Shukla}
\affil[1]{FLOW Research Center, Department of Engineering Mechanics, KTH Royal Institute of Technology, Stockholm, SE-10044, Sweden.}
\affil[2]{PDC Center for High Performance Computing, KTH Royal Institute of Technology, Stockholm, SE-11428, Sweden}
\affil[3]{Division of Applied Mathematics, Brown University, Providence, RI 02906, USA}

\maketitle

\begin{abstract}
The goal of this work is to investigate the capability of a neural operator (DeepONet) to accurately capture the complex deformation of a platelet's membrane under shear flow. The surrogate model approximated by the neural operator predicts the deformed membrane configuration based on its initial configuration and the shear stress exerted by the blood flow. The training dataset is derived from particle dynamics simulations implemented in LAMMPS. The neural operator captures the dynamics of the membrane particles with a mode error distribution of approximately 0.5\%. The proposed implementation serves as a scalable approach to integrate sub-platelet dynamics into multi-scale computational models of thrombosis.
\end{abstract}

\section{Introduction}
\label{sec:intro}
To accurately describe the complex and hierarchical dynamical phenomena of the human cardiovascular system, mathematical models capable of spanning multiple space-time scales are essential~\cite{zhang2016multi}.
Thrombosis, the pathological formation of blood clots within a blood vessel, is no exception~\cite{xu2008multiscale}.
At the cellular level, the fundamental mechanism underpinning thrombosis is platelet activation~\cite{tomaiuolo2017regulation}.
This process marks the transition of a platelet from a resting state to an active state, enabling adhesion to other platelets and the vessel wall through morphological changes and the release of pro-coagulant factors~\cite{kamath2001platelet}.
Platelet activation is triggered by various factors, including vascular injury, shear stress, and biochemical signals~\cite{yun2016platelet}.
For example, when a blood vessel is damaged, exposed collagen and von Willebrand factor interact with platelets, initiating clot formation~\cite{kroll1991willebrand}.
High shear stress, particularly in areas of stenosis (i.e., narrowed blood vessels), can also activate platelets by inducing mechanical forces that enhance their adhesiveness~\cite{zhang2002platelet}.
Additionally, chemical signals such as ADP, thrombin, and thromboxane A2, released by damaged cells or other activated platelets, bind to specific platelet receptors to promote activation and aggregation~\cite{sang2021interplay}.
Each of these cellular activation pathways is profoundly influenced by the large-scale characteristics of blood flow, a dependence that is especially significant in high shear-induced activation, which is the focus of this study.

Due to the large spatio-temporal range of scales that bridge the microscopic sub-platelet description of platelet activation with the macroscopic continuous description of blood flow, achieving a comprehensive multi-scale description of thrombosis presents significant mathematical and computational challenges~\cite{gutierrez2024decoding}.
Consequently, the two main approaches are based on either computational fluid dynamics (CFD)~\cite{hong2020evaluating} or Particle Dynamics (PD)~\cite{wang2023multiscale}, which focus on the macroscopic (blood flow) and microscopic (single platelet) aspects of thrombosis, respectively.

CFD methods operates under the assumption that blood can be modeled as a continuum medium flowing within either simplified~\cite{laudato2024sound,laudato2024analysis,laudato2023buckling} or patient-specific segmented geometries~\cite{karmonik2008computational,sundstrom2023machine,bornemann2024relation}.
The primary objective is to identify regions in the domain characterized by high shear stress and other macroscopic descriptors relevant to thrombogenesis~\cite{celestin2015computational,rorro2024backflow,fiusco2023effect}.
Additionally, the one-way interaction between blood flow and platelets can be explored using multi-phase flow methods, which treat platelets as suspended particles transported by the blood flow governed by the Navier-Stokes equations~\cite{bouchnita2021multiphase}.
This approach allows platelet activation to be incorporated into the modeling framework.
However, these methods heavily depend on low-order models~\cite{nobili2008platelet}, which are often difficult to parametrize, resulting in inaccurate predictions in realistic and clinically relevant scenarios~\cite{fuchs2019modeling}.

PD methods focus on a granular description of individual platelets and other blood components, such as red blood cells.
The dynamics of a platelet and the surrounding shear flow are modeled as millions of particles interacting through specific potentials~\cite{filipovic2008modelling}.
These models effectively capture the internal dynamics and complex biochemistry underlying platelet activation~\cite{zhang2021predictive}, adhesion~\cite{zhang2014multiscale}, and aggregation~\cite{gupta2021multiscale}.
However, the simplified assumptions about blood flow, necessary for computational feasibility, limit the applicability of PD models to clinically relevant and complex cases.

The development of a new multi-scale modeling paradigm that combines the strengths of these two approaches while overcoming their computational bottlenecks would be a significant milestone in the description of thrombogenesis~\cite{schwarz2023beyond}.
Recent advancements in reliable scientific machine learning methods present a timely opportunity to achieve this goal~\cite{hu2024tackling,yin2021non}.
An interesting approach was proposed in~\cite{shankar2022three}, where an artificial neural network approximated the results of a reduced-order model for the coagulation cascade, embedded within a 3D CFD simulation limited to single platelet resolution.
The key advantage of such implementations is that, once trained, the surrogate models approximated by neural network-based methods can be integrated with a CFD solver at virtually zero computational cost.
However, the primary challenge lies in the limited generalization of the surrogate models, which often require retraining when applied to different geometries.

In this work, we explore the potential of a specific class of neural network architectures, known as neural operators \cite{lu2021learning}, to accurately capture the complex morphological transformations of a single platelet under shear flow.
This investigation serves as a foundational step toward developing a new multi-scale modeling paradigm for studying pathophysiological blood flows.
The core idea is to utilize a neural operator that integrates macroscopic boundary conditions from a CFD solver (e.g., local shear stress around a platelet) with input points from particle dynamics simulations to predict the activation state of a platelet. This approach enables bridging the microscopic sub-platelet resolution with the macroscopic space-time scales typical of clinical data.

In the present work, we focus on a scenario where a single platelet deforms under the influence of a linear shear flow (Couette flow).
Platelet deformation is modeled using dissipative particle dynamics simulations implemented in LAMMPS~\cite{thompson2022lammps}.
The primary objective is to develop a surrogate model for this dynamics using DeepONet~\cite{lu2021learning}.
The network accepts two inputs: (1) the shear stress around the platelet, encoded in the particle dynamics simulation through the slope of the Couette flow velocity profile, and (2) the platelet undeformed configuration, represented by the initial positions of its constituent particles.
The network output is the deformed configuration of the platelet after one Jeffery's orbit period~\cite{taylor1923motion}.
The extent to which the neural operator can accurately resolve the platelet microscopic dynamics sets the limits of what can be achieved with the multi-scale modeling framework in future applications.
The impact of the network architecture and the size of the input space containing training dataset on its performance is investigated.
In addition, the model extrapolation capability for the range of shear stresses used in training is analyzed.

The paper is organized as follows: in Sec.~\ref{sec:particle}, the details of the particle dynamics model for the platelet in a Couette flow are discussed. In Sec.~\ref{sec:learning}, the DeepONet implementation and the details of the training are presented. Finally, in Sec.~\ref{sec:sensitivity}, the results of the meta-parametric sensitivity study are discussed.

\section{Particle dynamics model of a platelet in shear flow}
\label{sec:particle}
The particle dynamics simulation aims to determine the deformed configuration of a simplified platelet model under the influence of a viscous shear flow. 
The blood flow around the platelet is modeled using the Dissipative Particle Dynamics (DPD) method~\cite{groot1997dissipative}.
The platelet membrane, initially shaped as an ellipsoid and consisting of approximately 18,000 particles, is held together by inter-particle harmonic bonds.
Fluid-platelet interactions are mediated by a force field that allows the platelet to flip and deform within the viscous flow.

One of the objectives of this study is to assess whether the neural operator can accurately predict the deformed configuration of the platelet membrane under varying shear flows.
To achieve this, a training dataset was constructed using 101 particle dynamics simulations, each characterized by a different slope of the blood flow velocity profile.
Shear values range from 50 to \qty{250}{Pa}, corresponding to conditions relevant for high-shear-induced platelet activation~\cite{akins1995results}.

\begin{figure}[h!]
    \centering
    \includegraphics[width=0.85\linewidth]{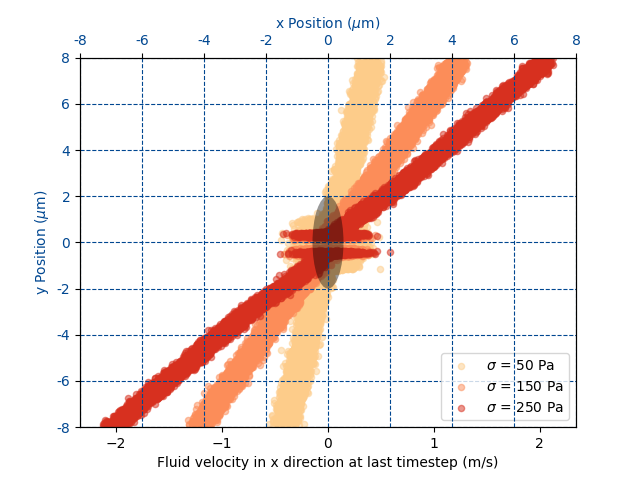}
    \caption{Diagram of the particle dynamics simulation domain (cross-section at $z=0$). The simulation box has dimensions 16x16x\qty{8}{\micro\meter}. The initial position of the platelet is shown as a gray ellipse. A Couette flow is generated by imposing the blood flow velocity at the boundaries. The resulting velocity profiles for shear stress values of $\sigma=50$~Pa, $\sigma=150$~Pa, and $\sigma=250$~Pa are shown by plotting the velocity of each blood particle against its position along the y axis. The velocity dispersion arises from the statistical temperature, while the horizontal lines near the platelet indicate the effects of non-penetration boundary conditions, which prevent blood particles from entering the platelet's membrane.}
    \label{fig:simulation_domain}
\end{figure}

The simulations are implemented in LAMMPS.
The solver employed is the standard velocity-Verlet algorithm, widely used in molecular dynamics simulations~\cite{groot1997dissipative}.
The simulation domain is a box with a height of \qty{16}{\micro\meter}, a length of \qty{16}{\micro\meter}, and a width of \qty{8}{\micro\meter} (see Fig.~\ref{fig:simulation_domain}).
The upper and lower boundaries of the domain are composed of ghost particles to enforce the no-slip boundary condition, while periodic boundary conditions are applied to the remaining boundaries.
The upper and lower walls (see Fig.~\ref{fig:simulation_domain}) move in opposite directions.
The velocity $U$ of the walls determines the shear stress $\sigma$ experienced by the platelet during its dynamics.
Each of the 101 different cases in the training dataset is characterized by a different shear stress and, in turn, by a different velocity of the walls, which are related by

\begin{equation}
    \label{eq:CouetteBcs}
    U=\sigma\frac{L}{2\mu}\,,
\end{equation}

\noindent

where $L=\qty{16}{\micro\meter}$ and $\mu$ is the dynamic viscosity of blood (see Tab.~\ref{tab:lammps}).

\begin{table}[h!]
    \centering
    \begin{tabular}{|l|c|}
        \hline
         Quantity & SI value \\
         \hline
         Blood viscosity & $9.84\times10^{-4}$ Pa s \\
         Blood particle mass & $1.10\times10^{-17}$ s\\
         Blood number density & $4.24\times10^{21}$ m$^{-3}$\\
         DPD cutoff radius & $1.60\times10^{-7}$ m \\
         DPD Temperature & $310.15$ K\\
         DPD Friction factor & $1.87\times10^{-11}$ Ns/m\\
         Platelet particle mass & $2.50\times10^{-19}$ kg\\
         Platelet elastic constant & $3\times10^{-3}$ N/m\\
         LJ cutoff radius & $1.42\times10^{-7}$ m\\
         LJ energy parameter & $6.85\times10^{-21}$ J \\
         LJ length parameter & $5.33\times10^{-8}$ m\\
         Time step & $2.40\times10^{-10}$ s \\
         \hline
    \end{tabular}
    \caption{LAMMPS parameters definition in SI units.}
    \label{tab:lammps}
\end{table}

The simulation time step (see Tab.~\ref{tab:lammps}) is selected to ensure good convergence of the solution and to accurately capture the complex deformation patterns of the platelet membrane. The total simulated physical time corresponds to one period of a Jeffery's orbit, which describes the trajectory of a rigid ellipsoid suspended in a shear flow. Since the platelet in this study is deformable, its final configuration differs from its initial one, hence this choice of simulation time provides a meaningful test for the ability of the neural operator to capture the platelet complex deformation dynamics.

\noindent
The details of the LAMMPS implementation for these simulations are outlined in the following subsections.

\subsection{The viscous shear flow model}
\label{ssec:flow}
Dissipative particle dynamics (DPD) is a well-established method to simulate the effects of macroscopic viscous flows at molecular level~\cite{zhang2014multiscale}.
In the case under analysis, the blood flow is represented in terms of effective particles interacting with each other according to the following relation~\cite{groot1997dissipative}:

\begin{equation}
    \label{eq:dpd}
    d\bm{v}_i = \frac{1}{m}\sum^N_{j=1}\left(\bm{F}_c dt+\bm{F}_d dt+ \bm{F}_r\sqrt{dt}\right)\,,
\end{equation}

\noindent
where $\bm{F}_c$, $\bm{F}_d$, and $\bm{F}_r$ represent the conservative, dissipative, and random forces, respectively.
The conservative force is defined as a function of the inter-particle distance $r_{ij}$ as $\bm{F}_c=\hat\alpha\omega_c(\bm{r}_{ij})\bm{e}_{ij}$, where 

\begin{equation}
    \omega_c (r_{ij})= \begin{cases}
        (1-r_{ij}/r_c)\,, r_{ij}<r_c\\
        0\,, r_{ij}>r_c
    \end{cases}
\end{equation}

\noindent
and $r_c$ is the cutoff radius of the interaction.
The dissipative force $\bm{F}_d$ depends on the velocity difference between two particles $\bm{v}_{ij}=\bm{v}_i-\bm{v}_j$ as $\bm{F}_d=-\hat\gamma\omega_d(r_{ij})(\bm{e}_{ij}\cdot\bm{v}_{ij})\bm{e}_{ij}$.
Finally, the random force is defined as $\bm{F}=\hat\sigma\omega_r(r_{ij})\zeta_{ij}\sqrt{dt}\bm{e}_{ij}$, where $\zeta_{ij}$ is a random Gaussian number with zero mean and unit variance.
The parameter $\hat\alpha$ represents the maximum inter-particle repulsion and it is defined as $\hat\alpha=75k_BT/\rho_fr_c$~\cite{groot1997dissipative}, where $\rho_f$ is the number density of the particle, $T$ is the temperature, and $k_B$ is the Boltzmann constant.
To ensure proper thermal equilibrium state, Espa\~nol et al.~\cite{espanol1995statistical} derived the relationship between the parameters $\hat\sigma$ and $\hat\gamma$ as $\hat\sigma^2=2\hat\gamma k_BT$.
Moreover, the relationship between the weight functions is $\omega_d=\omega_r^2$.
Consequently, the only free parameters left are the friction factor $\hat\gamma$ and the cutoff radius $r_c$ which have been estimated and validated by Bluestein et al.~\cite{zhang2014multiscale} for a similar experimental setup.
All the numerical values of the parameters can be found in Table~\ref{tab:lammps}.

\subsection{Blood-platelet interaction}
\label{ssec:fsi}
The platelet is modeled as a hollow ellipsoid of dimensions 4x4x\qty{1}{\micro\meter} made of approximately 18000 spring-connected particles located at the nodes of a network of triangles created using 3D Delaunay triangulation. The harmonic bond energy is defined as

\begin{equation}
    \label{eq:bonds}
    U_{harm} = \sum_{\text{bonds}}K(r-r_0)^2\,,
\end{equation}

\noindent
where $K$ is the elastic constant (see Tab.~\ref{tab:lammps}) and $r_0$ is the distance between two particles in the initial configuration. The value of the elastic constant has been estimated following~\cite{zhang2014multiscale} to ensure that the global elastic parameters (Young module and Poisson ratio) of the platelet are compatible with experimental measurements~\cite{haga1998quantification}. The platelet center is initially located at the intersection of the symmetry plans of the domain (see Fig.~\ref{fig:simulation_domain}).

\begin{figure}[h!]
    \centering
    \includegraphics[width=0.9\linewidth]{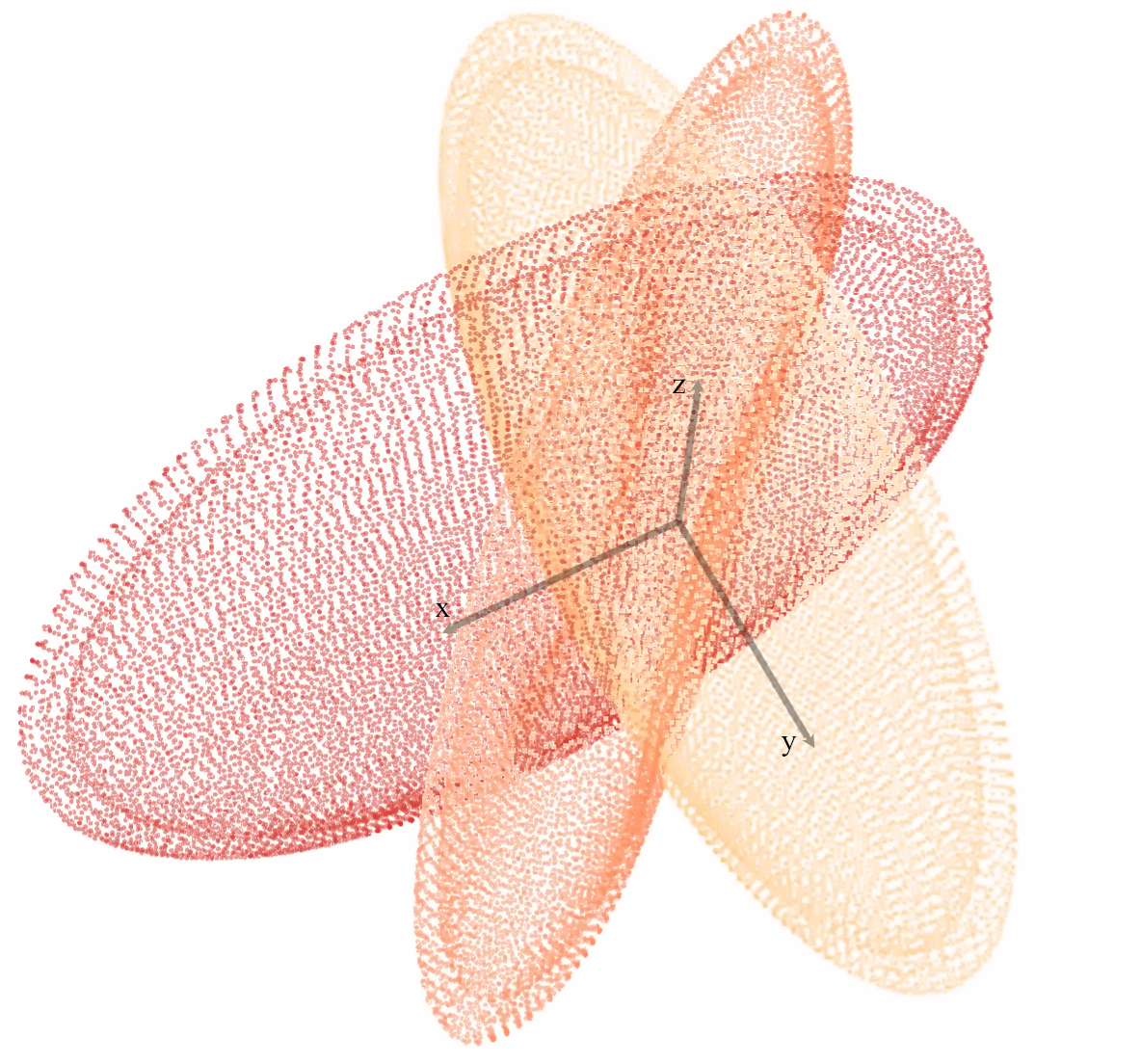}
    \caption{Comparison of the initial (yellow) and final configuration of the platelet subjected to Couette flow for shear stress values of $\sigma=50$~Pa (orange) and of $\sigma=250$~Pa (red) obtained from particle dynamics simulations using LAMMPS. Each of the unit vectors representing the x, y, and z axes is \qty{1}{\micro\meter} long.}
    \label{fig:scatter3d_init_last}
\end{figure}

The interaction between the blood flow introduced in Sec.~\ref{ssec:flow} and the platelet is mediated via a non-bonded pair-wise interaction, defined as

\begin{equation}
    \label{eq:fsi}
    d\bm{v}_i=\frac{1}{m}\sum^N_{j\neq i}\left(\nabla U(r_{ij})dt+F_d dt+F_r\sqrt{dt}\right)\,,
\end{equation}

\noindent
where $U$ is the Lennard-Jones potential

\begin{equation}
    \label{eq:LJ}
    \nabla U(r_{ij}) = 4\varepsilon_{LJ}\left[\left(\frac{\sigma_{LJ}}{r_{ij}}\right)^{12}-\left(\frac{\sigma_{LJ}}{r_{ij}}\right)^6\right],
\end{equation}

\noindent
which approximates both attractive and repulsive forces between non-bonding particles. The dissipative and random forces in Eq.~(\ref{eq:fsi}) allow to exchange mechanical and thermodynamical momentum between the platelet and the blood, while the drag-repulsive interaction allows to achieve no-slip boundary condition on the surface of the platelet. All the parameters in Eq.~(\ref{eq:LJ}) are listed in Tab.~\ref{tab:lammps} and have been validated in~\cite{zhang2014multiscale}. In Fig.~\ref{fig:scatter3d_init_last}, three snapshots of the platelet configuration are shown: one corresponding to the initial configuration of the platelet, two corresponding to its final configuration for the maximum and the minimum shear stress values.

\section{Learning the platelet dynamics}
\label{sec:learning}
The neural network architecture used to model the dynamics of the platelet under varying shear flow conditions is DeepONet, a deep operator network designed to approximate nonlinear continuous operators, as supported by the universal approximation theorem~\cite{lu2021learning}.
The operator $\bm{\xi} = \bm{\xi}(\bm{\xi_0}, T; \sigma)$, approximated by DeepONet, represents the displacement field $\bm{\xi}$ of each particle in the platelet initial configuration $\bm{\xi_0}$, given the shear stress $\sigma$, and evaluated after one Jeffery's orbit period, $T$.

\subsection{DeepONet implementation}
\label{ssec:don}
\begin{figure}[h!]
    \centering
    \begin{minipage}{0.495\textwidth}
        \centering
        \includegraphics[width=\textwidth]{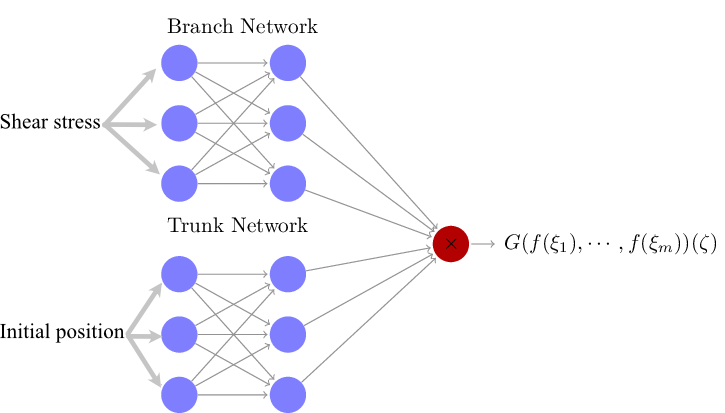}
        \caption{Schematics of the general architecture of DeepONet. The branch net encodes the boundary and initial conditions and other parameters of the PDEs. The trunk net encodes the independent variables at which the output functions are evaluated. The two outputs are combined performing a inner product.}
        \label{fig:don}
    \end{minipage}
    \hfill
    \begin{minipage}{0.495\textwidth}
        \centering
        \includegraphics[width=\textwidth]{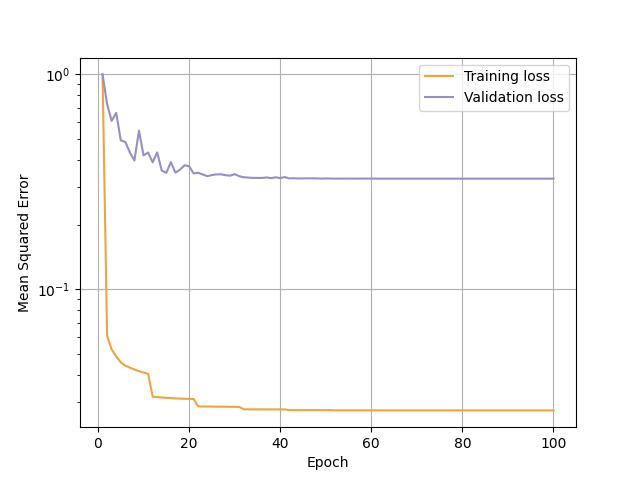}
        \caption{Training and validation loss functions, calculated as the mean squared error between the network predictions and the ground truth. The validation dataset represents 20\% of the total training dataset.}
        \label{fig:loss}
    \end{minipage}
\end{figure}

DeepONet typically comprises two deep neural networks (see Fig.~\ref{fig:don}): the branch network, which encodes the boundary and initial conditions as well as the parameters of the PDEs, and the trunk network, which encodes the independent variables at which the output functions are to be evaluated. In the specific implementation used in this work, both the branch net and the trunk net are designed as fully connected feed-forward neural networks (FCNs). The branch net takes as input the scalar value of the shear stress $\sigma$, while the trunk net takes as input the vectors representing the coordinates of the initial configuration of the platelet's particles. The outputs of the branch and trunk nets are then combined by performing an inner product between the outputs of branch and trunk networks and subsequently passed through a dense layer with three output dimensions.

The rectified linear unit (ReLU) activation function is used for both FCNs, while a linear activation function is applied in the output layer to enable the prediction of both positive and negative values.
The loss function employed is the mean squared error (MSE) between the deformed platelet configuration predicted by the network and that obtained from the corresponding particle dynamics simulation.
As shown in Fig.~\ref{fig:loss}, the training converges after 100 epochs.
The neural operator is implemented in Python using the TensorFlow library~\cite{tensorflow2015-whitepaper}.

\subsection{Prediction of platelet deformation} 
\label{ssec:predictions}
\begin{figure}[ht]
    \centering
    \begin{minipage}{0.495\textwidth}
        \centering
        \includegraphics[width=\textwidth]{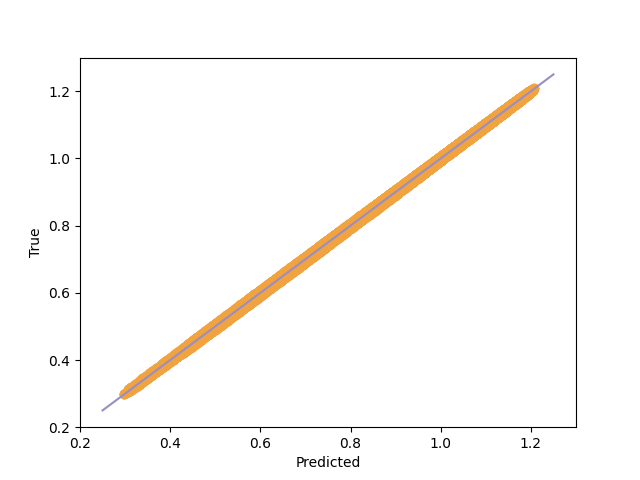}
        \caption{Parity plot for the DeepONet implementation. The diagonal line represents the perfect match between the neural operator prediction and the ground truth.}
        \label{fig:parity}
    \end{minipage}
    \hfill
    \begin{minipage}{0.495\textwidth}
        \centering
        \includegraphics[width=\textwidth]{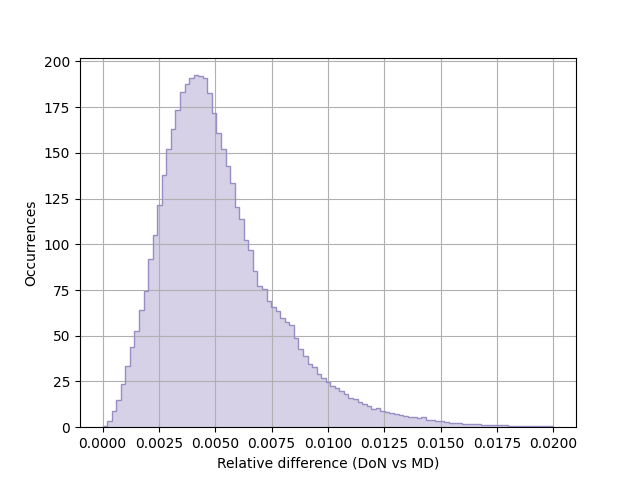}
        \caption{Distribution of the relative differences between the neural operator prediction and the ground truth. The maximum relative difference is 2\%, while the mode of the distribution is approximately 0.5\%.}
        \label{fig:histo_baseline}
    \end{minipage}
\end{figure}

\begin{figure}[h!]
    \centering
    \includegraphics[width=0.8\linewidth]{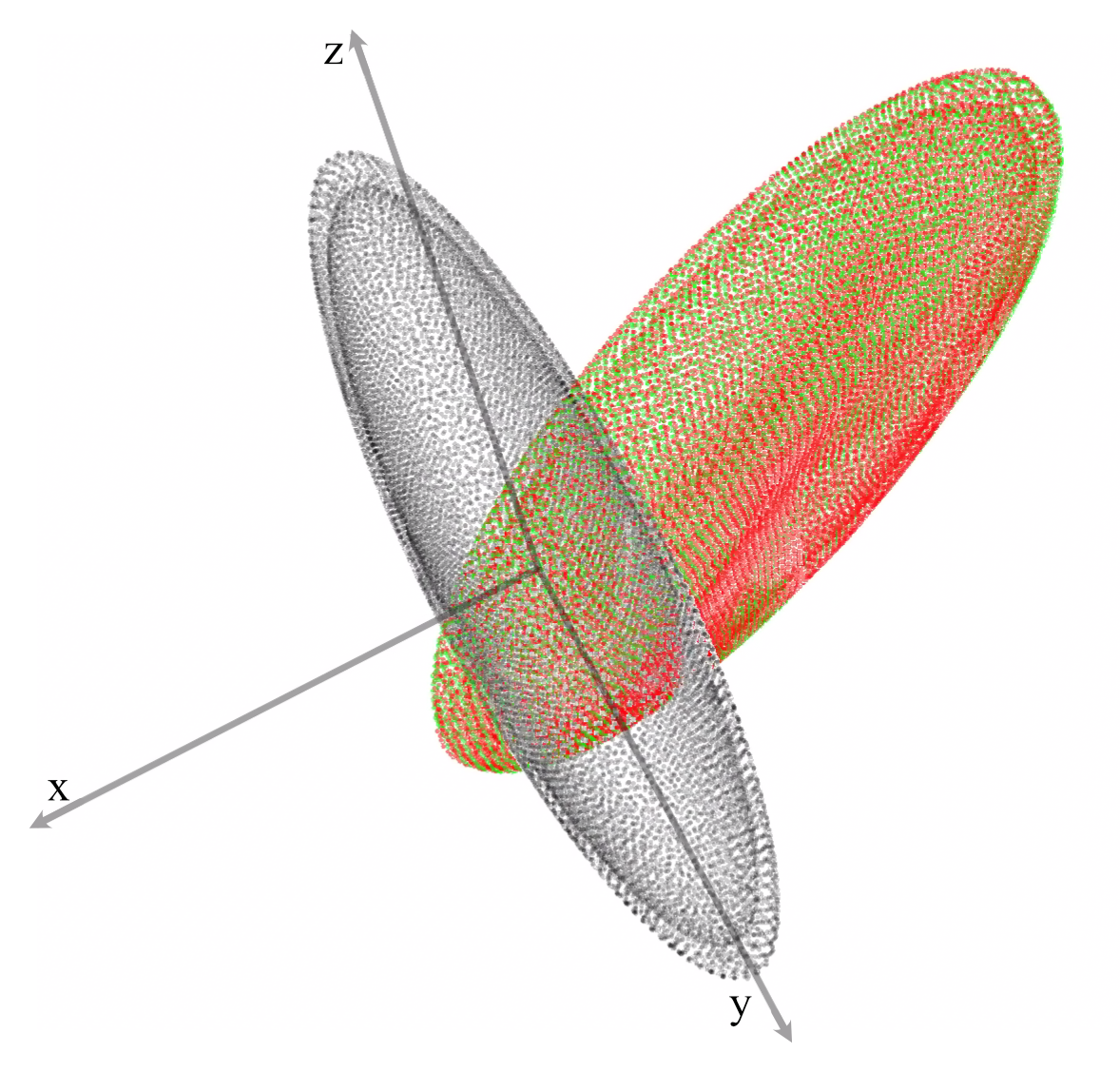}
    \caption{Comparison of the deformed platelet configuration predicted by DeepONet and that obtained from particle dynamics simulations using LAMMPS. The figure displays the displacement of approximately 18,000 particles, with green points representing the DeepONet predictions and red points indicating the ground truth simulation results. The gray points correspond to the platelet initial configuration.}
    \label{fig:DonVsLammps}
\end{figure}

The proposed DeepONet implementation effectively reproduces the results of the particle dynamics simulations.
The specific architecture was selected based on the meta-parameter sensitivity studies discussed in Sec.~\ref{sec:sensitivity}.
Once trained, DeepONet accurately predicts the deformed configuration of a platelet subjected to a given shear flow after one Jeffery's orbit.
Fig.~\ref{fig:parity} presents a parity plot comparing the ground truth and predicted values, where the diagonal line represents a perfect match.
Additional details on the relative difference distribution between the predicted displacements and the ground truth are shown in Fig.~\ref{fig:histo_baseline}.
The maximum relative difference is less than 2\%, with the mode being approximately 0.5\%.
One example of deformed platelet configurations predicted by DeepOnet is shown in Fig.~\ref{fig:DonVsLammps} alongside its respective ground truth configuration.
The neural operator model demonstrates a high degree of accuracy in predicting both the global orientation of the platelet in the shear flow and the geometric features of the deformed platelet membrane.

\section{Sensitivity study}
\label{sec:sensitivity}
The sensitivity studies presented in this section are motivated by the need to bridge the predictive capabilities of the proposed DeepONet implementation with clinically relevant data.
In patient-specific scenarios, clinical measurements such as the minimum and maximum shear stress derived from 4D flow MRI, provide critical input for modeling platelet activation dynamics.
The main objective of these studies is to establish how the neural operator framework can be effectively tailored to accommodate such data while ensuring accurate predictions, efficient training, and robust performance under varying conditions.
The section is structured into three subsections, each addressing a specific aspect of sensitivity analysis critical for integrating clinical data into the modeling workflow. 

In subsection \ref{ssec:size} the required size of the network is analyzed in terms of the number of parameters needed to solve the specific cases presented in this study.
While this aspect does not directly link to clinical data, minimizing network size is essential to reduce computational costs during model training, a practical consideration for deploying the framework in a clinical setting.

Subsection \ref{ssec:interpolation} focuses on determining the minimum number of particle dynamics simulations needed to train the network.
For example, given clinical measurements of minimum and maximum shear stress within a patient-specific geometry, this study aims to quantify how sparsely the training dataset can be sampled while maintaining accuracy.
This analysis is crucial for optimizing the computational cost associated with generating training data from particle dynamics simulations.

Finally, subsection \ref{ssec:extrapolation} explores the robustness of the neural network to uncertainties in clinical data, particularly variations in the measured shear stress values. It investigates how far the model can extrapolate beyond the range of shear stresses used in training while still maintaining a predefined maximum error. This study provides insights into the reliability of the model when faced with imprecise or uncertain input data.

\subsection{Network size}
\label{ssec:size}
The first study focuses on identifying an efficient architecture for the DeepONet implementation.
Here, efficiency is defined as the use of the smallest number of model parameters that achieves a maximum relative difference of less than 2\% between the network predictions and the ground truth.
This relative difference is calculated as the magnitude of the difference between the predicted and true displacement vectors of the platelet:

\begin{equation}
    \label{eq:rel_diff}
    D=\frac{||\xi_{pred}-\xi_{true}||}{||\xi_{true}||}
\end{equation}

\noindent
The difference is evaluated on a test dataset that was not used during the training of DeepONet.

\begin{figure}[h!]
    \centering
    \includegraphics[width=0.65\linewidth]{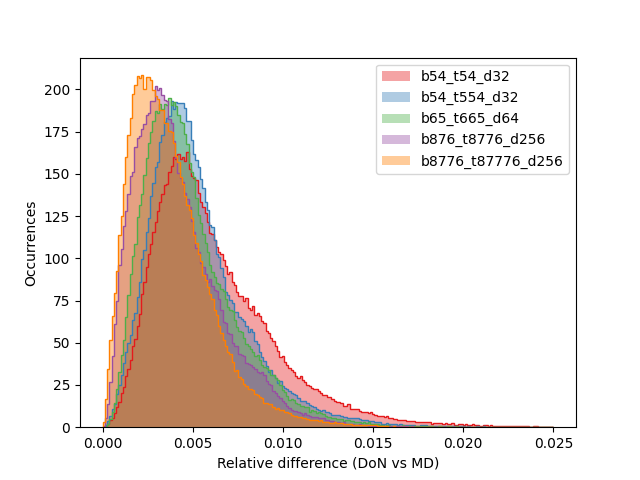}
    \caption{Relative difference distribution between the DeepONet predictions and the ground truth.
        The colors indicate different architectures.
        The selected architecture, labeled as \textit{b54\_t554\_d32}, consists of 2 layers in the branch net (one with $2^5=32$ and one with $2^4=16$ neurons) and 3 in the trunk net (two with $2^5=32$ and one with $2^4=16$ neurons), and a latent dimension of 32 neurons, totaling 6467 trainable parameters.
        This architecture achieves a maximum relative difference below 2\% and a mode of approximately 0.5\%.
        Larger architectures, labeled using the same convention, do not show significant improvements. All the curves have normalized total area equal to one.}
    \label{fig:size}
\end{figure}

Various architectures were tested by varying the depth and width of the trunk net, the branch net, and the latent space dimension. 
The results, shown in Fig.~\ref{fig:size}, illustrate the relative difference between the ground truth and the network predictions.
Notably, even for the smallest model tested, the mode of the difference distribution is approximately 0.5\%.
The second smallest architecture, consisting of 6467 trainable parameters, achieves a maximum difference below 2\% and a distribution mode of approximately 0.5\%.
More complex architectures do not demonstrate significant improvements.

As a result, this architecture will be used for the subsequent analyses and have been employed to produce the results in Sec.~\ref{sec:learning}. 

\subsection{Sparsity of the training data} 
\label{ssec:interpolation}
The second analysis focuses on determining the minimum number of shear stress examples required in the training dataset to achieve a maximum relative difference below 2\%, as defined in Eq.~(\ref{eq:rel_diff}).
This information is crucial for reducing the computational cost associated with training the network.

\begin{figure}[h!]
    \centering
    \includegraphics[width=0.65\linewidth]{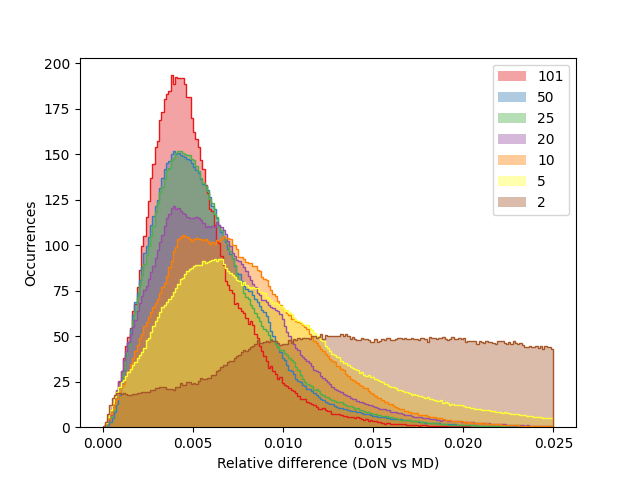}
        \caption{Relative difference distribution for different training dataset sizes, where the labels represents the number $N$ of uniformly spaced shear stress examples included in the training dataset.
        The relative difference is computed as defined in Eq.~(\ref{eq:rel_diff}).
        The results indicate that increasing $N$ beyond 25 does not significantly improve accuracy, suggesting that $N=25$ is sufficient to train the network while maintaining the desired maximum relative difference below 2\%.}
    \label{fig:interpolation}
\end{figure}

The study was conducted by testing 7 different training datasets, each used to train the DeepONet architecture identified in the previous subsection.
In each case, the training dataset comprises $N$ uniformly spaced shear stress examples.
The cases range from $N=2$, including only the minimum and maximum shear-induced platelet displacements, to $N=101$, representing the complete training dataset.
The relative difference, shown in Fig.~\ref{fig:interpolation}, was computed using Eq.~(\ref{eq:rel_diff}) with the excluded cases.

The results indicate that cases with $N\geq25$ do not show significant improvement, suggesting that $N=25$ is sufficient to train the network to the desired accuracy.
It is important to note that this sensitivity study focuses solely on the portion of the training dataset relevant to the branch net.
A separate analysis investigating the network’s ability to achieve similar accuracy by varying the number of particles constituting the platelet's membrane (relevant to the trunk net) will be the subject of a follow-up study.

\subsection{Extrapolation}
\label{ssec:extrapolation}
The final study investigates the ability of the DeepONet implementation identified in Sec.~\ref{ssec:size} to extrapolate beyond the shear stress examples provided during the training phase.
This capability is crucial for achieving robustness against uncertainties in clinical data, particularly in determining the extreme shear stress values in a given clinical scenario.

The study evaluates the relative difference, as defined in Eq.~(\ref{eq:rel_diff}), for three scenarios where progressively larger portions of the extreme shear stress examples are excluded from the training dataset.
In the first scenario, the first and last 10\% of shear stress examples are omitted.
In the second scenario, 20\% of the extremes are excluded, and in the third, 30\%.
In all the cases, the training dataset comprises 25 equally spaced shear stress examples, following the results discussed in the previous section.
This approach assesses the network's ability to extrapolate and maintain accuracy when faced with shear stress values outside the training range.

\begin{figure}[h!]
    \centering
    \includegraphics[width=0.65\linewidth]{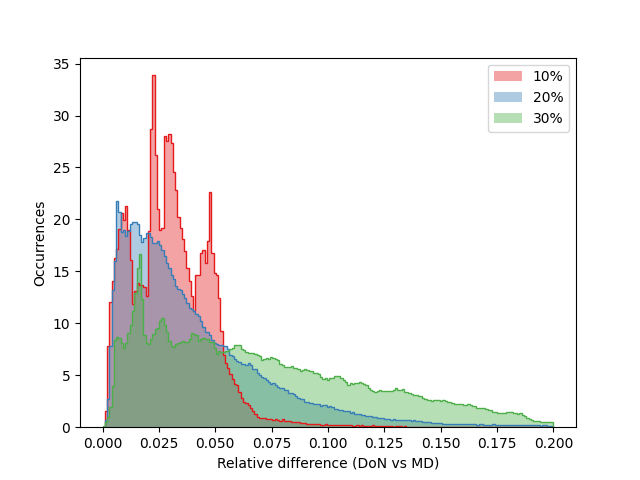}
        \caption{Relative difference distribution for the three extrapolation scenarios, where progressively larger portions of extreme shear stress examples are excluded from training.
        The scenarios correspond to the omission of the first and last 10\%, 20\%, and 30\% of shear stress examples.
        The relative difference is computed as defined in Eq.~(\ref{eq:rel_diff}).
        The first scenario shows a maximum relative difference below 10\%, with peaks in the difference distribution between 2.5\% and 5\%.
        The second and third scenarios result in larger differences, exceeding 20\%, highlighting the challenges of extrapolation beyond the training range.}
    \label{fig:extrapolation}
\end{figure}

\begin{figure}[h!]
    \centering
    \includegraphics[width=0.8\linewidth]{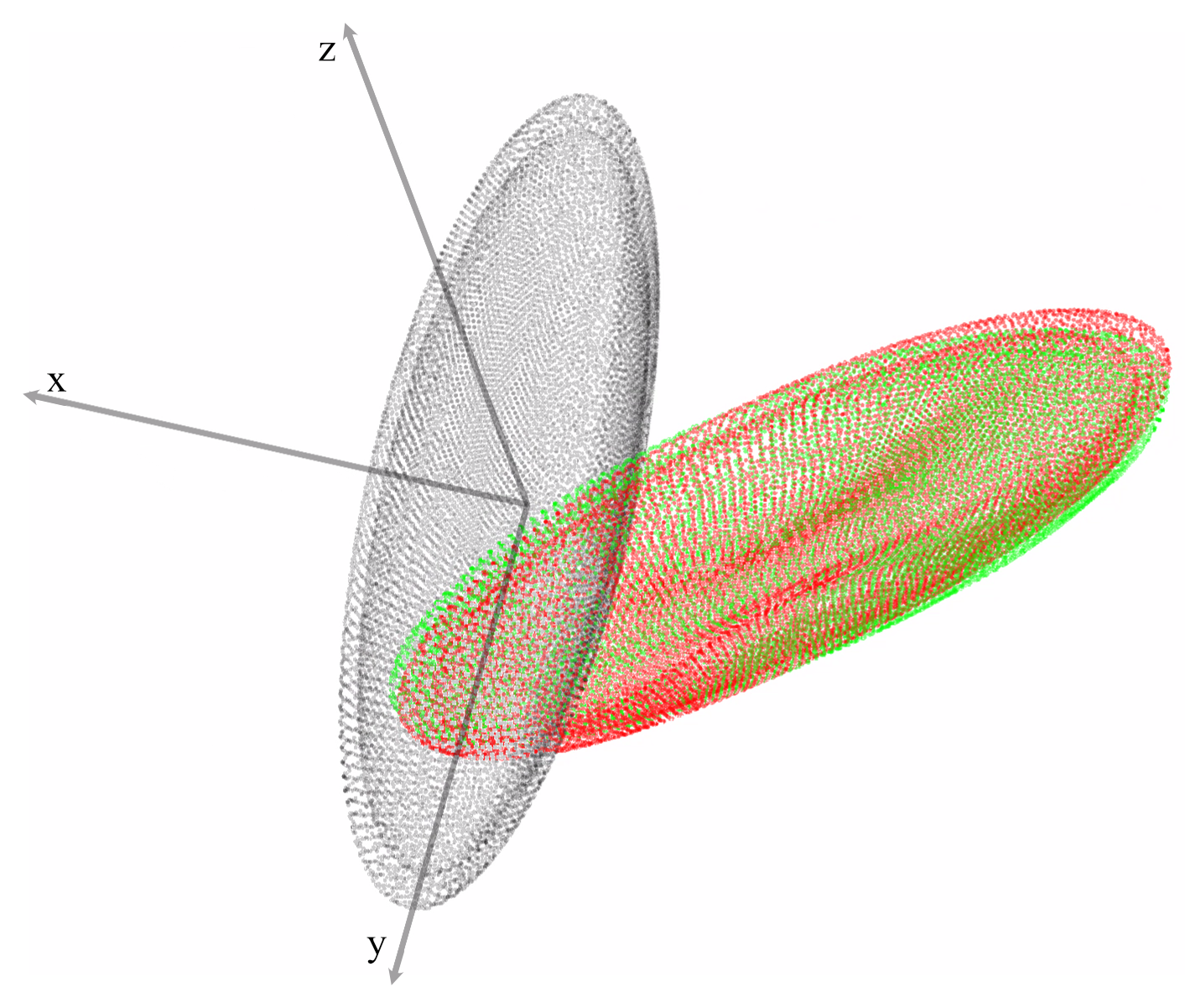}
    \caption{Comparison of the deformed platelet configuration predicted by DeepONet (green points) and that obtained from particle dynamics simulations using LAMMPS (red points). The shear stress in this case lies outside the training dataset, representing an extrapolation of approximately 5\% beyond the training range.}
    \label{fig:scatter_extrapolation}
\end{figure}

The results, shown in Fig.~\ref{fig:extrapolation}, indicate that in the first scenario, the DeepONet implementation can extrapolate with a maximum relative difference below 10\%. 
Peaks in the difference distribution are observed in the range of 2.5\% to 5\%.
One instance of the corresponding predicted platelet displacement is compared with the ground truth in Fig.~\ref{fig:scatter_extrapolation}.
More ambitious extrapolations, such as the second and third scenarios, result in larger relative differences, exceeding 20\%.

\section{Conclusions}
\label{sec:conclusions}
The results of this work demonstrate the capability of DeepONet to accurately capture the complex dynamics of shear-induced deformation as modeled by dissipative particle dynamics simulations.
The relatively small number of parameters required to achieve a relative difference below 2\% highlights the potential of this approach to extend to more intricate microscopic descriptions of platelet activation, aggregation, and adhesion.
These promising findings pave the way for future investigations.

One interesting perspective is to train the network to model the entire dynamical evolution of a platelet by incorporating time as a parameter in the trunk net.
Another direction involves enhancing the particle dynamics models to include the platelet cytoskeleton and cytoplasm, enabling the study of more realistic platelet activation pathways.
Increasing the complexity of the platelet description would allow for a more comprehensive understanding of platelet activation mechanisms.

These efforts need to be framed within the broader context of multi-scale thrombosis modeling.
The ability of neural operators like DeepONet to capture the microscopic dynamics of individual platelets, as shown in this study, provides a foundation for scaling this approach to simulations involving thousands of platelets in macroscopic multi-phase fluid dynamics models.
The virtually zero computational cost of obtaining sub-platelet scale information via DeepONet can be leveraged to bridge the gap between microscopic particle dynamics and macroscopic fluid dynamics simulations at clinically relevant scales.
In this context, the present study represents a significant first step toward integrating microscopic and macroscopic scales in thrombosis modeling.

\section*{Acknowledgement}
M.L. was supported by Swedish Research Council Grant No. 2022–03032.
The numerical computations were enabled by resources provided by the National Academic Infrastructure for Supercomputing in Sweden (NAISS) at the PDC Center for High Performance Computing, KTH Royal Institute of Technology, Sweden, partially funded by the Swedish Research Council through grant agreement no. 2022-06725.
The authors want to thank prof. George Karniadakis for the fruitful discussions and the scientific support.

\section*{Competing Interests}
The authors have no conflicts of interest to declare that are relevant to the content of this chapter.


\end{document}